\def   \ni {\noindent}

\def   \ssk {\vskip  5truept}

\def   \bsk {\vskip 15truept}

\def   \newline {\hfil\break}

\def \gray     {$\gamma$-ray }
\def \grays    {$\gamma$-rays }

\def \sig      {$\sigma$ }

\def\fwb{62mm}

\def\fwh{52mm}

\documentstyle[epsfig]{article}

\begin{document}

\hsize 5truein
\vsize 8truein
\font\abstract=cmr8
\font\keywords=cmr8
\font\caption=cmr8
\font\references=cmr8
\font\text=cmr10
\font\affiliation=cmssi10
\font\author=cmss10
\font\mc=cmss8
\font\title=cmssbx10 scaled\magstep2
\font\alcit=cmti7 scaled\magstephalf
\font\alcin=cmr6 
\font\ita=cmti8
\font\mma=cmr8
\def\ref{\par\noindent\hangindent 15pt}
\null


\title{\ni COMPTEL OBSERVATIONS OF AGN AT MEV-ENERGIES 
}                                               

\bsk \bsk
\author{\ni W. Collmar$^{1}$, K. Bennett$^{4}$, H. Bloemen$^{2}$,
J.J. Blom$^{5}$, W. Hermsen$^{2}$, G.G. Lichti$^{1}$, J.~Ryan$^{3}$,
V. Sch\"{o}nfelder$^{1}$, J.G. Stacy$^{3}$, H. Steinle$^{1}$, 
O.R. Williams$^{4}$, C. Winkler$^{4}$
}                                                       
\bsk

\affiliation{1)  Max-Planck-Institut f\"ur extraterrestrische Physik, 
D-85740 Garching, Germany}
    
\affiliation{2) SRON-Utrecht, Sorbonnelaan 2, 3584 CA Utrecht, The Netherlands}

\affiliation{3) University of New Hampshire, IEOS, Durham NH 03824, USA}

\affiliation{4) Astrophysics Division, SSD/ESA, NL-2200 AG Noordwijk, The Netherlands}

\affiliation{5) INAOE, Apartado Postal 216 y 51, 72000 Puebla, Pue, M\'exico}

\bsk
\baselineskip = 12pt

\abstract{ABSTRACT \ni The COMPTEL experiment aboard CGRO, exploring the previously unknown sky at MeV-energies, has so far detected 10 Active Galactic Nuclei (AGN): 9 blazars and the radio galaxy Centaurus~A. No Seyfert galaxy has been found yet. With these results COMPTEL has opened the field of
extragalactic \gray astronomy in the MeV-band.   
 }                                                    
\bsk
\baselineskip = 12pt

\bsk
\baselineskip = 12pt


\text{\ni 1. SOURCE DETECTIONS
\ssk
\ni     

Before the launch of the Compton Gamma-Ray Observatory (CGRO) the 'MeV-sky', 
where the COMPTEL experiment (0.75 - 30 MeV) operates, was hardly explored. Prior to COMPTEL three Active Galactic Nuclei (AGN) had been reported to emit detectable \grays at these energies: two Seyfert galaxies and the radio galaxy Centaurus~A. In particular, no quasar was known to emit in this energy band.

\begin{table}[hbt]
\caption{Table 1. List of COMPTEL-detected AGN for which a dedicated analysis has been done.}
\vspace*{-0.4cm}
\begin{center}\begin{tabular}{lccccccc}
\hline
Source & z & Det.$^1$ & Energy$^2$ & Spectral & O/E$^3$  & Type$^4$   & Lum.$^5$ \\
ID            &       &      & Bands  & Shape       & Det. & FS,AT,BT & [10$^{47}$ erg/s]    \\
\hline
PKS 0208-512  & 1.003 & m    & yyyy   & ---         & ny  & yQM      &  45 \\
GRO J0516-609 & 1.09  & s    & nyyn   & soft        & yy  & yQM      &  48 \\
PKS 0528+134  & 2.06  & m    & yyyy   & 1.9$\pm$0.4 & yy  & yQE      & 151 \\
PKS 1222+216  & 0.435 & s    & nnyn   & ---         & yy  & yQE      & 1.4 \\
3C 273        & 0.158 & m    & yyyy   & 2.0$\pm$0.4 & yy  & yQE      & 0.9 \\
3C 279        & 0.538 & m    & yyyy   & 1.9$\pm$0.4 & yy  & yQE      & 4.5 \\
Cen A        & 0.0007 & m    & yyyy   & 2.3$\pm$0.1 & y?  & nR.      & 8.7 $\times$10$^{41}$ \\
PKS 1622-297  & 0.815 & s    & nnyy   & hard        & yy  & yQE      & 16 \\
CTA 102       & 1.037 & s    & nnny   & ---         & yy  & yQE      & 22 \\
3C 454.3      & 0.859 & s    & nnny   & ---        & yy  & yQE      & 17 \\
\hline
\end{tabular}\end{center}
\footnotesize
further source indications (no detailed analysis): PKS~0446+112 (Kuiper et al. 1996)
\newline
$^1$Detections: s - single, m - multiple; $^3$OSSE/EGRET detection
\newline
$^2$Detection in the 4 standard COMPTEL energy bands: y - detected,
 n - not detected 
\newline
$^4$Source type: FS: flat radio spectrum; AT (AGN type): Q - quasar, R - radio galaxy; BT (blazar type):
 E - EGRET type, M - MeV blazar
\newline
$^5$time-averaged (if possible), isotropic MeV-luminosity for the energy band of detection  
\normalsize
\end{table}

Soon after the launch of CGRO, EGRET, also aboard CGRO and observing at energies above 30~MeV, reported the detection of several radio-loud AGN, quasars or BL Lacertae objects, belonging to the so-called blazar subclass of AGN. 
Currently nine of the $\sim$70 EGRET blazars
(e.g. Hartman et al. 1997) have been detected by COMPTEL in one or more of its four standard energy bands (0.75-1~MeV, 1-3~MeV, 3-10~MeV, 10-30~MeV). So far COMPTEL has detected only flat-spectrum radio quasars; no BL Lac-type blazar has been seen yet. These sources are often visible during flaring events reported by EGRET. In many cases, the detections occur near threshold, which indicates that COMPTEL is sensitive to the strongest MeV-sources only. Apart from blazars only one further AGN, the radio galaxy Centaurus~A, has been seen so far, leading to a total of 10 AGN detections.
These sources are listed in Table~1 and some are visible in the skymap
of Fig.~1.   

Prior to launch Seyfert galaxies had been promising candidates for MeV-emission. However, despite an extensive Seyfert search in the COMPTEL data,
 no object has been found yet (e.g. Maisack et al. 1995).
This is consistent with the findings by the OSSE experiment, also aboard CGRO, 
that their hard X-ray spectra cut off at energies around $\sim$100~keV (e.g. Johnson et al. 1997).

\bsk
\ni 2. SOURCE PROPERTIES 
\ssk
\ni

\begin{figure} [tb] 
 \centerline{\psfig{file=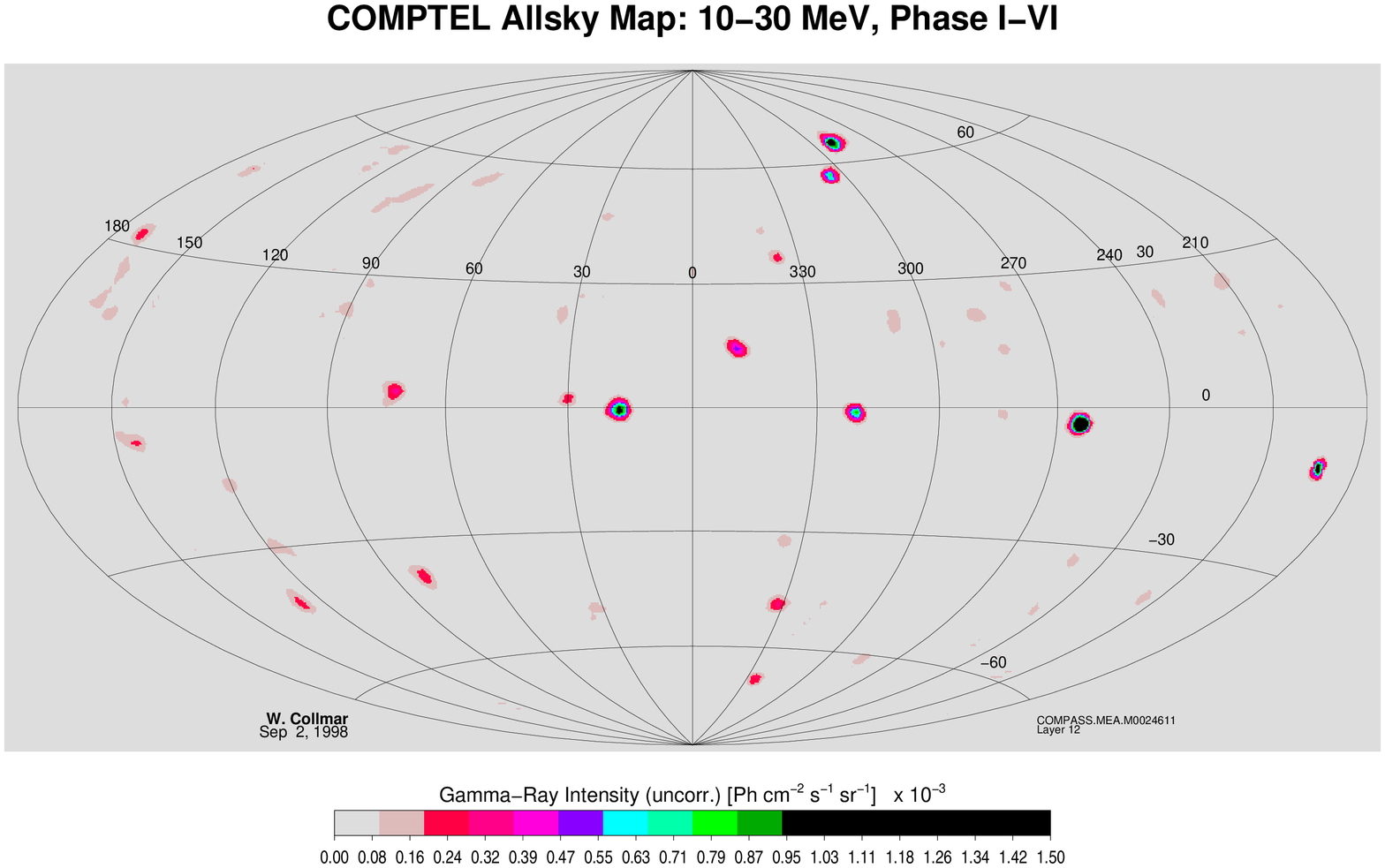,width=5.5cm,height=12.0cm,angle=90.,clip= } } 
\caption{FIGURE 1.
 COMPTEL Phase I to VI (April '91 to November '97) all-sky flux map in the 10-30~MeV band. The Galactic diffuse emission as well as the emission from the Crab has been subtracted. Evidence for several AGN is clearly visible:
3C~273 (l/b: 290/64), 3C~279 (l/b: 305/57), PKS~0528+134 (l/b: 191/-11), 
and PKS~1622-297 (l/b: 349/13).
 }
\end{figure}


\begin{figure} [bt]
 \begin{picture}(125,51)(0,0)
    \put(0,0){%
       \makebox(60,0)[lb]{\psfig{file=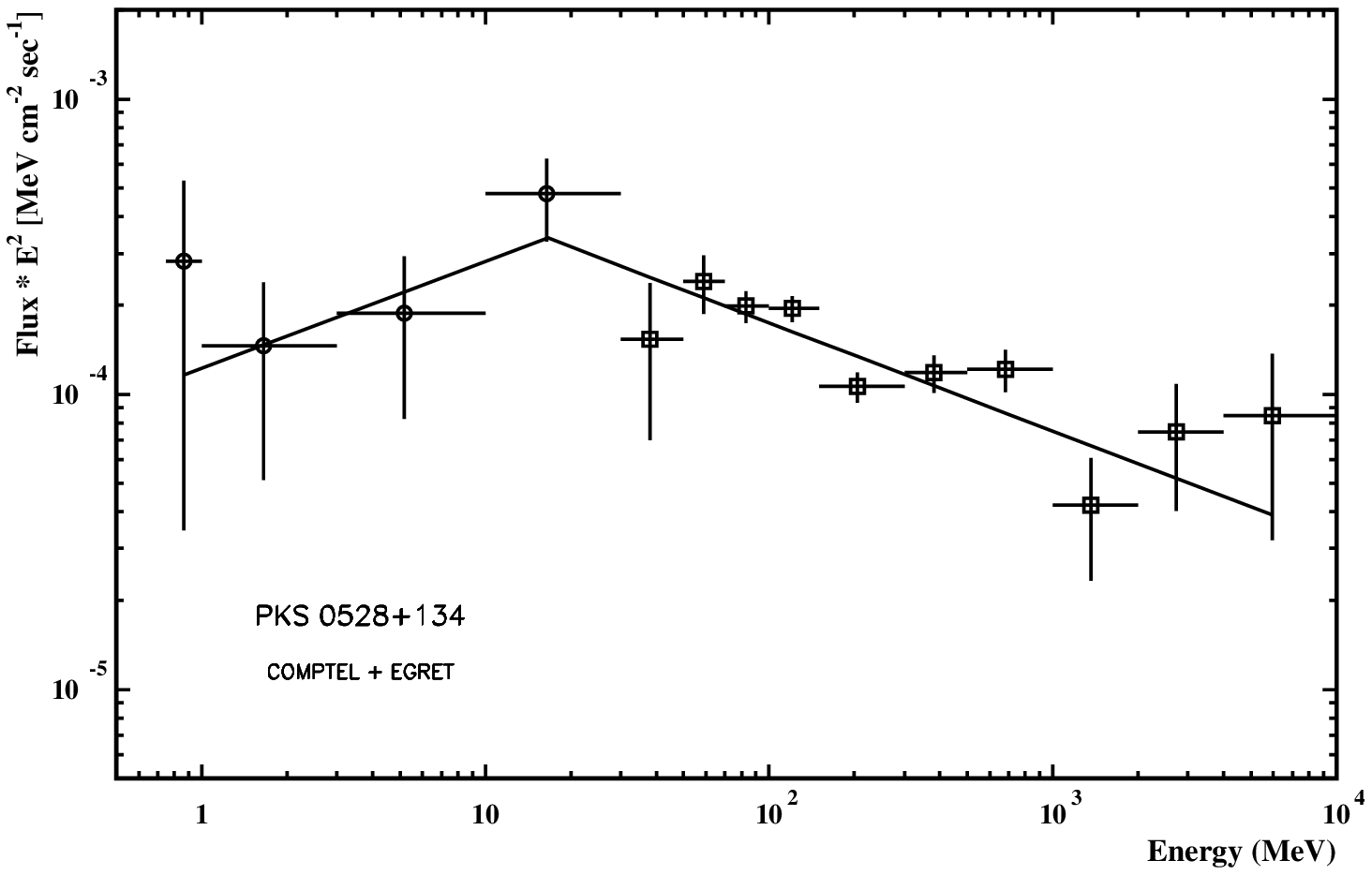,width=\fwb,height=\fwh,clip=}}}
    \put(65,0){%
       \makebox(60,0)[lb]{\psfig{file=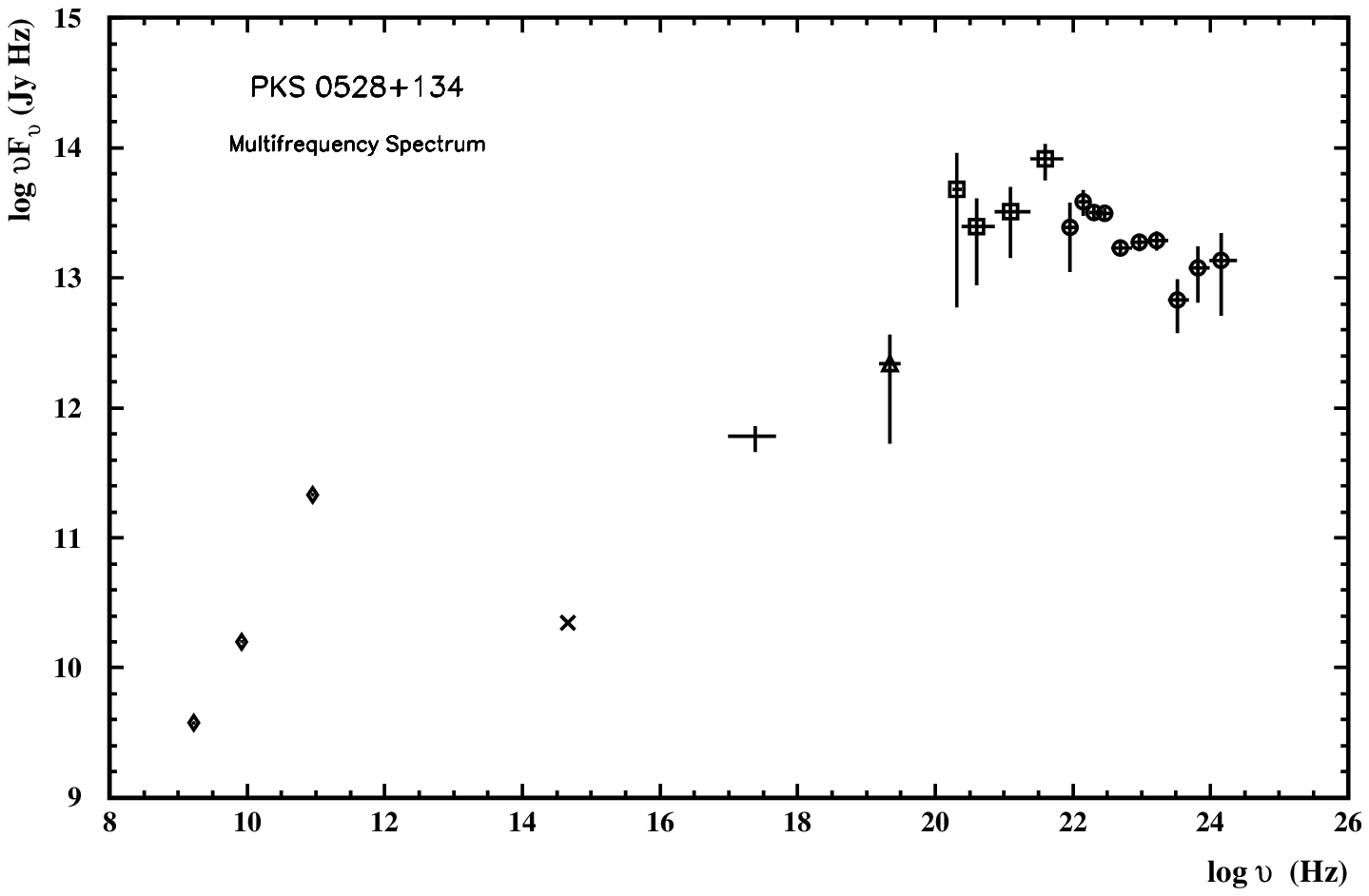,width=\fwb,height=\fwh,clip=}}}
\end{picture}
\caption{FIGURE 2. 
Left:
 Simultaneous COMPTEL ($\circ$) and EGRET ($\Box$) spectrum of PKS~0528+134 in April 1991. The spectral turnover at MeV-energies is evident.
Right:
Non-simultaneous broad-band spectrum for the same observational period. The luminosity peaks at MeV-energies and the bolometric luminosity is dominated by $\gamma$-rays. \label{fig2} }
\end{figure}

No AGN is each time visible when it is located within the COMPTEL field-of-view. This indicates time variability of their MeV-emission of the order of months to years. For the strongest sources 
(e.g. 3C~273) this variability is statistically significant. Five sources - 3C~273, 3C~279, PKS~0528+134, PKS~0208-512, and Cen~A - have been detected several times throughout the mission. 
The blazar 3C~273, which also was the prime AGN candidate before launch, has been seen most often. It is most significantly detected in the 1-10~MeV band
(Collmar et al. 1996) and exceeds the 3\sig detection threshold 
in $\sim$80\% of all individual Virgo pointings.
The shortest time variability was observed for 3C~279. During a flaring period early in 1996 it showed a flux increase by a factor of $\sim$4 in the 10-30~MeV band within 10 days (Collmar et al. 1997a), in close agreement with the variations as observed by EGRET above 100~MeV.   
The other five sources have only been detected during certain time periods,
like 
PKS~1622-297 during a four-week \gray flaring period (Collmar et al. 1997a) and 3C~354.3 and CTA~102 in a combination of CGRO viewing periods during the early CGRO mission (Blom et al. 1995a). 

Because the sources are often detected near threshold in only one or two of the COMPTEL bands, the knowledge on their MeV-spectra is limited.  Nevertheless, some trends are apparent. Four sources - 3C~273,
3C~279, PKS~0528+134, and Cen~A - have been seen significantly enough to estimate their spectral shape at MeV-energies. In time-averaged analyses the spectra are well described by power-law shapes (E$^{-\alpha}$) with a
photon index $\alpha$ of the order of 2. During flare periods at energies above 100~MeV, the MeV-shapes are usually harder ($\alpha <$ 2). This indicates that mainly the high-energy ($>$3~MeV) part of the COMPTEL band is following this flux increase and therefore may suggests an additional high-energy component which emerges during such flaring periods. The spectral variability of PKS~0528+134 for example, fits into this picture (Collmar et al. 1997b). 
The average Cen~A spectrum is softer than the typical blazar spectrum and
connects nicely to the hard X-ray spectrum measured by OSSE (Steinle et al. 1998).           

Combining COMPTEL AGN spectra with spectral results from neighboring energy bands (OSSE, EGRET) shows that for COMPTEL-detected blazars a spectral turnover occurs within, or close to, the COMPTEL band (Fig.~2). This is confirmed by multiwavelength analyses which, however, also show, that during \gray flares the luminosity across the whole electromagnetic spectrum peaks near the COMPTEL band (e.g. 3C~279, PKS~0528+134), and that the bolometric luminosity is dominated by \gray emission. The MeV-luminosities of blazars are typically between 10$^{47}$
and 10$^{49}$ erg/s, if one assumes isotropic emission. 
The most luminous source is PKS~0528+134, which reaches values larger than
10$^{49}$erg/s.

In addition to the regular EGRET-type blazars, COMPTEL has provided evidence for so-called 'MeV-blazars'. They are exceptionally bright in the 1-10~MeV range when compared to simultaneous fluxes measured with EGRET above 30~MeV. 
Two objects, GRO~J0506-609 (Bloemen et al. 1995) and PKS~0208-512 (Blom et al. 1995b), have been found to show this behaviour on occasions. 

\bsk
\ni 3. SUMMARY
\ssk
\ni 
After more than 7 years, COMPTEL has provided new, interesting, and also unexpected results on extragalactic \gray sources. Apart from the previously known radio galaxy Centaurus~A, COMPTEL detected nine, in this energy range previously unknown quasars, thereby opening the field of extragalactic \gray astronomy at MeV-energies.  
Because COMPTEL is still `healthy' and may stay in operation for further years, more and improved results on AGN can be expected in the future.

\bsk
\baselineskip = 12pt
{\abstract \ni ACKNOWLEDGMENTS
This research was supported by the Deutsche Agentur f\"ur Raumfahrtangelegenheiten (DARA) under the grant 50 QV 90968, by NASA under   contract NASA-26645, and by the Netherlands Organisation for Scientific Research.}

\bsk
\baselineskip = 12pt


{\references \ni REFERENCES
\ssk

\ref Bloemen, H., Bennett, K., Blom, J.J. et al. 1995, A\&A 293, L1
\ref Blom, J.J., Bloemen, H., Bennett, K. et al. 1995a, A\&A 295, 330
\ref Blom, J.J., Bennett, K., Bloemen, H. et al. 1995b, A\&A 298, L33
\ref Collmar, W., Bennett, K., Bloemen, H. et al. 1996, A\&AS 120, No.4, 515
\ref Collmar, W., Bennett, K., Bloemen, H. et al. 1997a, AIP 410, 1341
\ref Collmar, W., Bennett, K., Bloemen, H. et al. 1997b, A\&A 328, 33
\ref Hartman, R., Collmar, W., von Montigny, C., Dermer, C.D. 1997, AIP 410, 307
\ref Johnson, W.N., Zdziarski, A., Madejski, G. et al. 1997, AIP 410, 283
\ref Kuiper, L., Hermsen, W., Bennett, K. et al. 1996, A\&A Suppl. 120, 73
\ref Maisack, M., Collmar, W., Barr, P. et al. 1995, A\&A 298, 400
\ref Steinle, H., Bennett, K., Bloemen, H. et al. 1998, A\&A 330, 97
}                      

\end{document}